\documentclass[lettersize,journal]{IEEEtran}
\usepackage{amsmath,amsfonts}
\usepackage{algorithm}
\usepackage{array}
\usepackage[caption=false,font=normalsize,labelfont=sf,textfont=sf]{subfig}
\usepackage{textcomp}
\usepackage{stfloats}
\usepackage{url}
\usepackage{verbatim}
\usepackage{graphicx}
\usepackage{cite}
\usepackage{amssymb}
\usepackage{graphicx}
\usepackage{algpseudocode}
\usepackage{amsthm}
\usepackage{color}

\begin{document}

\title{Channel Estimation for Delay Alignment Modulation}

\author{Dingyang Ding and Yong Zeng,~\IEEEmembership{Senior Member,~IEEE}\vspace{-1cm}
\thanks{This work was supported by the National Key R\&D Program of China with
Grant number 2019YFB1803400.}
\thanks{The authors are with the National Mobile Communications Research
Laboratory, Southeast University, Nanjing 210096, China. Y. Zeng is also
with the Purple Mountain Laboratories, Nanjing 211111, China (e-mail: \{220200693, yong\_zeng\}@seu.edu.cn). (\emph{Corresponding author: Yong Zeng.)}}}



\maketitle

\begin{abstract}
Delay alignment modulation (DAM) is a promising technology for inter-symbol interference (ISI)-free communication without relying on sophisticated channel equalization or multi-carrier transmissions. The key ideas of DAM are \emph{delay pre-compensation} and \emph{path-based beamforming}, so that the multi-path signal components will arrive at the receiver simultaneously and constructively, rather than causing the detrimental ISI. However, the practical implementation of DAM requires channel state information (CSI) at the transmitter side. Therefore, in this letter, we study an efficient channel estimation method for DAM based on block orthogonal matching pursuit (BOMP) algorithm, by exploiting the block sparsity of the channel vector. Based on the imperfectly estimated CSI, the delay pre-compensations and tap-based beamforming are designed for DAM, and the resulting performance is studied. Simulation results demonstrate that with the BOMP-based channel estimation method, the CSI can be effectively acquired with low training overhead, and the performance of DAM based on estimated CSI is comparable to the ideal case with perfect CSI.
\end{abstract}


\section{Introduction}
Delay alignment modulation (DAM) is a promising technology to achieve inter-symbol interference (ISI)-free communications in time-dispersive channels, by exploiting the super spatial resolution  of large antenna arrays and multi-path sparsity of millimeter wave (mmWave) channels \cite{1}.  The key ideas of DAM are {\it delay pre-compensation} and {\it path-based beamforming}. Specifically, by introducing appropriate delays for the information-bearing symbols and applying path-based transmit beamforming, all the multi-path signal components may arrive simultaneously and constructively at the receiver, thus contributing to the enhancement of the desired signal, rather than causing detrimental ISI. This is achieved without relying on sophisticated conventional techniques like channel equalization or multi-carrier signal transmission. Compared to the dominate orthogonal frequency division multiplexing (OFDM) technology, DAM avoids the drawbacks like high peak-to-average-power ratio (PAPR), severe out-of-band (OOB) emission and vulnerability to carrier frequency offset (CFO) \cite{1}. Furthermore, DAM may also achieve higher spectral efficiency than OFDM as it requires less guard interval overhead \cite{1,2}. On the other hand, different from OFDM, DAM is effective mainly for spatially sparse channels with large antenna arrays, so that the significant multi-path signal components corresponding to different delays can be well resolved in the spatial domain. Besides, the implementation of DAM requires time-domain channel state information (CSI), i.e., the channel vectors of each delay tap.\par
Channel estimation has been extensively studied for wireless communication systems \cite{3,4,5,6,7,8,9,10,11,12,13}. For example, a joint pilot design and downlink channel estimation scheme based on deep learning for multiple-input multiple-output (MIMO)-OFDM was proposed in \cite{3}. In \cite{4}, the channel information is obtained by estimating the direction-of-arrival (DoA) and the channel gain. Hierarchical search based on two specially designed codebook was proposed to acquire the CSI in \cite{5}. Besides, channel impulse response (CIR) can be estimated based on the least square (LS) or linear minimum mean square error (LMMSE) algorithms \cite{6}. For channels with inherent sparsity property, various sparse recovery and compressive sensing techniques can be applied for CIR estimation \cite{7,8,9,10,11,12,13}. For example, in \cite{7}, a two-stage channel estimation method was proposed, which detects the non-zero taps before estimating the CIR. Besides, the matching pursuit (MP) algorithm \cite{8} and orthogonal matching pursuit (OMP) algorithm \cite{9} were adopted for CIR estimation by maximizing the correlation of a column of the dictionary matrix with the residual signal. Based on the joint sparsity of wideband mmWave channel in angular-delay domain, block orthogonal matching pursuit (BOMP) algorithm was applied to estimate the CIR for hybrid multiple-input multiple-output (MIMO) architecture \cite{10}.\par
However, most existing channel estimation schemes mentioned above are designed for multi-carrier OFDM systems, which are not directly applicable  to the new single-carrier DAM systems. Therefore, in this letter, we study the channel estimation scheme tailored for DAM and investigate the performance of DAM with imperfect CSI. Specifically, by exploiting uplink-downlink reciprocity,
the channel in downlink is estimated by uplink training. With multi-path sparsity, channel estimation can be reformulated as a block-sparse signal recovery problem, so that BOMP algorithm can be applied to efficiently acquire the channel by exploiting the block sparsity \cite{14}. Based on the estimated channel, the delay pre-compensations and tap-based beamforming are designed for DAM. Simulation results are provided to demonstrate the effectiveness of the BOMP-based channel estimation scheme, and show that DAM with estimated channel using small training overhead can achieve comparable performance as the ideal case with perfect CSI.
\section{System Model and DAM with Perfect CSI}
We consider a wideband multiple-input single-output (MISO) communication system, where the base station (BS) with $M\gg 1$ antennas wishes to communicate with a single-antenna user equipment (UE). The extension of DAM to the more general MIMO systems has been studied in \cite{15} and its channel estimation problem will be pursued in our future work. We assume quasi-static block fading environment, where the channel remains constant within each coherent block and may vary across different blocks. The $k$th delay tap of the downlink baseband channel can be expressed as
\begin{equation}\label{eq:1}
\textbf{h}_{\mathrm{DL}}^{H}[k]=\sum\nolimits_{l=1}^{L}\textbf{h}_{l}^{H}p(kT_{s}-\tau_{l}),k=0,...,K-1,
\end{equation}
where $L$ denotes the number of multi-paths, $p(t)$ is the pulse shaping function, $T_{s}$ is the sampling interval, $\textbf{h}_{l}\in\mathbb{C}^{M\times 1}$ and $\tau_{l}$ denote the channel vector and delay for the $l$th multi-path, respectively. Note that different from the existing DAM works \cite{1,2}  that use Kronecker delta function for time-domain channel modelling, we use the more practical model in \eqref{eq:1} by considering pulse shaping function $p(t)$. Furthermore, the multi-path delays $\tau_l$ do not have to be integer multiples of $T_s$. The total number of delay taps is $K=\lceil\tau_{\mathrm{UB}}/T_s\rceil$, where $\tau_{\mathrm{UB}}$ is  a sufficiently large delay
value beyond which no significant power can be received. For mmWave massive MIMO systems with large antenna arrays and multi-path spatial sparsity \cite{16,17}, we have $L\ll K$ and $L\ll M$. In this case, out of the $K$ channel taps, only around $L$ taps have significant power.\par
By exploiting the multi-path channel sparsity and super spatial dimension of mmWave large-array systems, a novel equalization-free single-carrier transmission technique called DAM was recently proposed based on Kronecker delta channel models \cite{1}. In this letter, we extend DAM for the more generic channel model in \eqref{eq:1}. To this end, we need to first find out the subset of $L^{\prime}$ most significant channel taps $\Omega=\{k_{1},...,k_{L^{\prime}}\}\subset \{0,...,,K-1\}$, e.g., by selecting channel taps with power no smaller than $C \mathop{\max}\limits_{k}\Vert\textbf{h}_{\mathrm{DL}}[k]\Vert^{2}$, where $C<1$ is certain threshold. Note that depending on the selected threshold $C$, $L^{\prime}$ may be smaller or slightly larger than the number of multi-paths $L$. Besides, the maximum delay in $\Omega$ is denoted as $k_{\mathrm{max}}=\mathop{\max}\limits_{1\leq l\leq L^{\prime}}k_{l}$. Then, the transmitted DAM signal for channel model (1) is
\begin{equation}\label{eq:2}
\textbf{x}[n]=\sum\nolimits_{l=1}^{L^{\prime}}\textbf{f}_{l}s[n-\kappa_{l}],
\end{equation}
where $s[n]$ denotes the information-bearing symbol sequence with normalized power  $\mathbb{E}[|s[n]|^{2}]=1$, $\textbf{f}_{l}\in\mathbb{C}^{M\times 1}$ and $\kappa_{l}=k_{\mathrm{max}}-k_{l}$ denote the per-tap transmit beamforming vector and the deliberately introduced delay pre-compensation for the $l$th significant delay tap in $\Omega$, respectively.
The received signal at the UE is
\begin{equation}\label{eq:3}
\begin{split}
y[n]&=\sum\nolimits_{k=0}^{K-1}\textbf{h}_{\mathrm{DL}}^{H}[k]\textbf{x}[n-k]+z[n]\\
&=\sum\nolimits_{k=0}^{K-1}\sum\nolimits_{l=1}^{L^{\prime}}\textbf{h}_{\mathrm{DL}}^{H}[k]\textbf{f}_{l}s[n-\kappa_{l}-k]+z[n]
\end{split}
\end{equation}
where $z[n]\sim\mathcal{CN}(0,\sigma^{2})$ is the additive white Gaussian noise (AWGN). The signal in \eqref{eq:3} can be equivalently written as
\begin{equation}\label{eq:4}
\begin{split}
y[n]&=\Big(\sum\nolimits_{l=1}^{L^{\prime}}\textbf{h}_{\mathrm{DL}}^{H}[k_{l}]\textbf{f}_{l}\Big)s[n-k_{\mathrm{max}}]+\\
&\sum\nolimits_{l=1}^{L^{\prime}}\sum\nolimits_{k\neq k_{l}}^{K-1}\textbf{h}_{\mathrm{DL}}^{H}[k]\textbf{f}_{l}s[n-k_{\mathrm{max}}+k_{l}-k]+z[n].
\end{split}
\end{equation}
When the UE is locked to the delay $k_{\mathrm{max}}$, the first term in \eqref{eq:4} is the desired signal while the second term is the ISI. Fortunately, when $M\gg L^{\prime}$, the ISI can be effectively mitigated by applying the simple tap-based beamforming, such as tap-based zero-forcing (ZF), maximum ratio transmission (MRT), and minimum mean-square error (MMSE) beamforming \cite{1}.
\par
{\it Tap-based ZF beamforming:} With tap-based ZF beamforming, the designed beamforming vectors $\{\textbf{f}_{l}\}_{l=1}^{L}$ satisfy
\begin{equation}\label{eq:5}
\textbf{h}^{H}_{\mathrm{DL}}[k_{l^{\prime}}]\textbf{f}_{l}=0, \forall l^{\prime}\neq l.
\end{equation}
As a result, the ISI term in \eqref{eq:4} associated with those significant channel taps in $\Omega$ can be eliminated. The ZF condition in \eqref{eq:5} is feasible almost surely as long as $M\geq L^{\prime}$. Let $\textbf{H}_{l}=[\textbf{h}_{\mathrm{DL}}[k_{1}],...\textbf{h}_{\mathrm{DL}}[k_{l-1}],\textbf{h}_{\mathrm{DL}}[k_{l+1}],...,\textbf{h}_{\mathrm{DL}}[k_{L^{\prime}}]]$, and $\textbf{Q}_{l}=\textbf{I}_{M}-\textbf{H}_{l}(\textbf{H}_{l}^{H}\textbf{H}_{l})^{-1}\textbf{H}_{l}^{H}$. Similar to \cite{1},  the ZF beamforming for DAM is
\begin{equation}\label{eq:6}
\textbf{f}^{\mathrm{ZF}}_{l}=\frac{\sqrt{P_{\mathrm{DL}}}\textbf{Q}_{l}\textbf{h}_{\mathrm{DL}}[k_{l}]}{\sqrt{\sum\nolimits_{l=1}^{L^{\prime}}\Vert\textbf{Q}_{l}\textbf{h}_{\mathrm{DL}}[k_{l}]\Vert^{2}}}, l=1,...,L^{\prime},
\end{equation}
where $P_{\mathrm{DL}}$ is the downlink transmit power.\par
{\it Tap-based MRT beamforming:}When $M<L^{\prime}$ so that ZF beamforming is infeasible, or when low-complexity implementation is desired, the MRT or MMSE beamforming can be applied. Similar to \cite{1}, the tap-based MRT beamforming is
\begin{equation}\label{eq:7}
\textbf{f}^{\mathrm{MRT}}_{l}=\frac{\sqrt{P_{\mathrm{DL}}}\textbf{h}_{\mathrm{DL}}[k_{l}]}{\Vert\textbf{h}_{\Sigma}\Vert}, l=1,\cdots,L^{\prime},
\end{equation}
where $\textbf{h}_{\Sigma}=[\textbf{h}_{\mathrm{DL}}^{T}[k_{1}],...,\textbf{h}_{\mathrm{DL}}^{T}[k_{L^{\prime}}]]^{T}\in\mathbb{C}^{ML^{\prime}\times1}$.\par
{\it Tap-based MMSE beamforming:}To obtain the optimal MMSE beamforming, we need to derive the signal-to-interference-plus-noise ratio (SINR) expression, by noting that different combinations of $k_{l}$ and $k$ of the ISI term in \eqref{eq:4} may correspond to identical symbols. As a result, we define the following effective channel to group those interfering symbols with identical delays \cite{1}
\begin{equation}\label{eq:8}
\textbf{g}^{H}_{l}[i]=
\begin{cases}
  \textbf{h}_{\mathrm{DL}}^{H}[k], & \mbox{if $\exists l\in\{1,...,L^{\prime}\}$, s.t. $k_{l}-k=i,$}\\
  \textbf{0}, & \mbox{otherwise},
\end{cases}
\end{equation}
where $i\in\{\pm1,\cdots,\pm (K-1)\}$.
Then, \eqref{eq:4} can be equivalently written as
\begin{equation}\label{eq:9}
\begin{split}
&y[n]=\Big(\sum\nolimits_{l=1}^{L^{\prime}}\textbf{h}_{\mathrm{DL}}^{H}[k_{l}]\textbf{f}_{l}\Big)s[n-k_{\mathrm{max}}]+\\
&\sum\nolimits_{i=-(K-1),i\neq0}^{K-1}\Big(\sum\nolimits_{l=1}^{L^{\prime}}\textbf{g}_{l}^{H}[i]\textbf{f}_{l}\Big)s[n-k_{\mathrm{max}}+i]+z[n].
\end{split}
\end{equation}
Hence, the SINR is
\begin{equation}\label{eq:10}
\begin{split}
\gamma&=\frac{|\sum\nolimits_{l=1}^{L^{\prime}}\textbf{h}_{\mathrm{DL}}^{H}[k_{l}]\textbf{f}_{l}|^{2}}{\sum\nolimits_{i=-(K-1),i\neq0}^{K-1}|\sum\nolimits_{l=1}^{L^{\prime}}\textbf{g}^{H}_{l}[i]\textbf{f}_{l}|^{2}+\sigma^{2}}\\
&=\frac{\textbf{f}_{\Sigma}^{H}\textbf{h}_{\Sigma}\textbf{h}_{\Sigma}^{H}\textbf{f}_{\Sigma}}{\textbf{f}_{\Sigma}^{H}(\sum\nolimits_{i=-(K-1), i\neq0}^{K-1}\textbf{g}_{\Sigma}[i]\textbf{g}_{\Sigma}^{H}[i]+\frac{\sigma^{2}}{\Vert\textbf{f}_{\Sigma}\Vert^{2}}\textbf{I})\textbf{f}_{\Sigma}},
\end{split}
\end{equation}
where $\textbf{f}_{\Sigma}=[\textbf{f}_{1}^{T}, \cdots, \textbf{f}_{L^{\prime}}^{T}]^{T}$ and $\textbf{g}_{\Sigma}[i]=[\textbf{g}_{1}^{T}[i], \cdots, \textbf{g}_{L^{\prime}}^{T}[i]]^{T}$.
Based on \eqref{eq:10}, the optimal tap-based MMSE transmit beamforming is
\begin{equation}\label{eq:11}
\textbf{f}_{\Sigma}^{\mathrm{MMSE}}=\sqrt{P_{\mathrm{DL}}}\frac{\textbf{C}^{-1}\textbf{h}_{\Sigma}}{\Vert\textbf{C}^{-1}\textbf{h}_{\Sigma}\Vert},
\end{equation}
where $\textbf{C}=\sum\nolimits_{i=-(K-1), i\neq0}^{K-1}\textbf{g}_{\Sigma}[i]\textbf{g}_{\Sigma}^{H}[i]+\frac{\sigma^{2}}{P_{\mathrm{DL}}}\textbf{I}$.\par
The DAM delay pre-compensation and tap-based beamforming designs discussed above assume that the BS has perfect knowledge of the channel of each delay tap, i.e., $\textbf{h}_{\mathrm{DL}}[k], k=0,...,K-1$. In practice, such CSI needs to be acquired via channel estimation. In the following, we study an efficient BOMP-based CSI estimation method for DAM via uplink training, by exploiting uplink-downlink reciprocity.
\section{BOMP-based Channel Estimation for DAM}
With channel reciprocity, the uplink channel is the transpose of the downlink channel, given by
\begin{equation}\label{eq:12}
\textbf{h}_{\mathrm{UL}}[k]=\textbf{h}_{\mathrm{DL}}^{T}[k]=\sum\nolimits_{l=1}^{L}\textbf{h}^{\ast}_{l}p(kT_{s}-\tau_{l}), k=0,..,K-1
\end{equation}
Note that to implement the DAM in Section II, it is sufficient to estimate the composite channel vectors $\textbf{h}_{\mathrm{DL}}[k]$ of each tap, without having to estimate the individual path delays $\tau_l$ and its channel vector $\textbf{h}_l$. For systems with multi-path sparsity, we have $K\gg L$.
With uplink-based channel estimation, during each coherent block, a pilot sequence of length $N$ is sent by the UE for channel estimation, which is denoted as $x[n], n=0,1,...,N-1$. The received training signal at the BS is
\begin{equation}\label{eq:15}
\textbf{y}[n]=\sqrt{P_{\mathrm{UL}}}\sum\nolimits_{k=0}^{K-1}\textbf{h}_{\mathrm{UL}}[k]x[n-k]+\textbf{z}[n],n=0,...,N-1
\end{equation}
where $P_{\mathrm{UL}}$ denotes the uplink transmit power, and $\textbf{z}[n]\in\mathbb{C}^{M\times1}\sim\mathcal{CN}(\textbf{0},\sigma^{2}\textbf{I}_{M})$ denotes the AWGN. Note that due to channel time-dispersion, the transmitted symbols in \eqref{eq:15} involve symbols $x[n]$ for $n<0$, which may be obtained from previous decodings in a data stream \cite{18}.\par
By stacking the $N$ received signal vectors into $\textbf{Y}=[\textbf{y}[0],...,\textbf{y}[N-1]]\in\mathbb{C}^{M\times N}$, \eqref{eq:15} can be compactly written in matrix form as
\begin{equation}\label{eq:16}
\textbf{Y}=\textbf{HX}+\textbf{Z},
\end{equation}
where
\begin{equation}\label{eq:17}
\textbf{X}=
\sqrt{P_{\mathrm{UL}}}
\begin{pmatrix}
x[0]&x[1]&\cdots&x[N-1]\\
x[-1]&x[0]&\cdots&x[N-2]\\
\vdots&\ddots&\vdots\\
x[1-K]&x[2-K]&\cdots&x[N-K]
\end{pmatrix},
\end{equation}
\begin{equation}\label{eq:18}
\textbf{H}=[\textbf{h}_{\mathrm{UL}}[0],...,\textbf{h}_{\mathrm{UL}}[K-1]], \textbf{Z}=[\textbf{z}[0],...,\textbf{z}[N-1]].
\end{equation}
Based on the received signal matrix $\textbf{Y}$ and the known pilot matrix $\textbf{X}$, the channel matrix $\textbf{H}$ can be estimated at the BS.\par
Note that the channel matrix $\textbf{H}$ to be estimated contains $MK$ unknown elements, and the received signal matrix $\textbf{Y}$ has $MN$ elements. Therefore, in general, at least $N\geq K$ training intervals is required for meaningful linear estimation. Fortunately, with multi-path sparsity, there are approximately $L\ll K$ significant channel taps. Therefore, the channel matrix $\textbf{H}$ exhibits column sparsity, which can be utilized to substantially reduce the training overhead.
With vectorization operation, \eqref{eq:16} can be written as
\begin{equation}\label{eq:19}
\textbf{c}=\textbf{A}\textbf{d}+\textbf{b},
\end{equation}
where $\textbf{c}=\mathrm{vec}(\textbf{Y}), \textbf{A}=(\textbf{X}^{T}\otimes\textbf{I}_{M}), \textbf{d}=\mathrm{vec}(\textbf{H})$, and $\textbf{b}=\mathrm{vec}(\textbf{Z})$. Note that $\textbf{c}\in\mathbb{C}^{MN\times1}$ is the measurement vector, $\textbf{A}\in\mathbb{C}^{MN\times MK}$ is the dictionary matrix with each column regarded as an atom, and $\textbf{d}\in\mathbb{C}^{MK\times1}$ is the block sparse vector to be estimated under the noise $\textbf{b}\in\mathbb{C}^{MN\times1}$. Therefore, estimating the channel matrix $\textbf{H}$ is equivalent to determining the values of significant blocks in $\textbf{d}$ based on the knowledge of $\textbf{c}$ and $\textbf{A}$, which can be solved effectively with compressed sensing techniques \cite{10,11,12}.

Specifically, by noting that the channel estimation process in \eqref{eq:19} is essentially a block sparse signal recovery problem, we use the BOMP algorithm \cite{14} to estimate $\textbf{d}$ efficiently.
For convenience, $\textbf{A}$ in \eqref{eq:19} can be expressed as a concatenation of column-blocks $\textbf{A}_{k}$ of size $MN\times M:$
\begin{equation}\label{eq:20}
\textbf{A}=[\underbrace{\textbf{a}_{1}\cdots\textbf{a}_{M}}_{\textbf{A}_{1}} \underbrace{\textbf{a}_{M+1}\cdots\textbf{a}_{2M}}_{\textbf{A}_{2}}\cdots\underbrace{\textbf{a}_{M(K-1)+1}\cdots\textbf{a}_{MK}}_{\textbf{A}_{K}}].
\end{equation}
\par
Similar to OMP algorithm \cite{19}, the basic idea of the BOMP algorithm is to minimize the residual between the measurement and the linear combination of selected atoms with estimated coefficients \cite{14}. Algorithm 1 summarizes the procedures of the BOMP-based channel estimation for our considered problem. Note that $\underline{\textbf{A}}$ is obtained by normalizing each column of the dictionary matrix $\textbf{A}$. In step 3 of Algorithm 1, at each stage $q$, we select the block $\underline{\textbf{A}}_{k_{q}}$ that is most correlated to the residual $\textbf{r}_{q-1}$ among those unselected blocks in the normalized dictionary $\underline{\textbf{A}}$. In step 5, the residual $\textbf{r}_{q}$ is updated by subtracting the measurement $\textbf{c}$ its projection on the space spanned by columns of those selected dictionary blocks $\textbf{A}_{\Gamma_{q}}$. The process is repeated until the difference of the norm of the residual between two iterations becomes less than a predefined threshold $\epsilon$.\par
\begin{algorithm}
\caption{BOMP-based channel estimation}
\label{alg:Framwork}
\begin{algorithmic}[1]
\Require
Measurement vector $\textbf{c}$, dictionary matrix $\textbf{A}$, normalized dictionary matrix $\underline{\textbf{A}}$,
\State
Initialize: Residual vector $\textbf{r}_{0}=\textbf{c}$, support set $\Gamma_{0}=\varnothing$, iteration index $q=1$, threshold $\epsilon$;
\While{$\Vert\textbf{r}_{q}\Vert_{2}-\Vert\textbf{r}_{q-1}\Vert_{2} > \epsilon$}
\State $k_{q}=\arg\max\limits_{1\leq k\leq K}\Vert\underline{\textbf{A}}^{H}_{k}\textbf{r}_{q-1}\Vert_{2}$;
\State $\Gamma_{q}=\Gamma_{q-1}\cup k_{q}$;
\State $\textbf{r}_{q}=\textbf{c}-\textbf{A}_{\Gamma_{q}}(\textbf{A}_{\Gamma_{q}}^{H}\textbf{A}_{\Gamma_{q}})^{-1}\textbf{A}_{\Gamma_{q}}^{H}\textbf{c}$;
\State $q=q+1$;
\EndWhile
\Ensure
Estimated support set $\Gamma_{q-1}$, estimated CIR vector $\hat{\textbf{d}}=(\textbf{A}_{\Gamma_{q-1}}^{H}\textbf{A}_{\Gamma_{q-1}})^{-1}\textbf{A}_{\Gamma_{q-1}}^{H}\textbf{c}$.
\end{algorithmic}
\end{algorithm}
\section{DAM Based on Estimated CSI}
In this section, we investigate the DAM design and resulting performance based on the estimated channel. Due to channel estimation errors, the estimated downlink channel is in general different from the true channel. Let $\hat{\textbf{h}}_{\mathrm{DL}}[k],k=0,...,K-1$ denote the estimated downlink channel based on Algorithm 1.
Similar to Section II, based on the estimated channel, the set of significant taps is selected, denoted as $\hat{\Omega}=\{\hat{k}_{1},...,\hat{k}_{\hat{L}}\}$, and the maximum delay is $\hat{k}_{\mathrm{max}}=\mathop{\max}\limits_{1\leq l\leq \hat{L}}\hat{k}_{l}$. Then, the transmitted signal for DAM with estimated channel is
\begin{equation}\label{eq:22}
\hat{\textbf{x}}[n]=\sum\nolimits_{l=1}^{\hat{L}}\hat{\textbf{f}}_{l}s[n-\hat{\kappa}_{l}],
\end{equation}
where the tap-based beamforming vector $\hat{\textbf{f}}_{l}$ and  the introduced delay $\hat{\kappa}_{l}$ are designed based on estimated channel.\par
The delay pre-compensation is set as $\hat{\kappa}_{l}=\hat{k}_{\mathrm{max}}-\hat{k}_{l},l=1,...,\hat{L}$. Furthermore, the path-based ZF and MRT beamforming are designed as
\begin{equation}\label{eq:23}
\hat{\textbf{f}}^{\mathrm{ZF}}_{l}=\frac{\sqrt{P_{\mathrm{DL}}}\hat{\textbf{Q}}_{l}\hat{\textbf{h}}_{\mathrm{DL}}[\hat{k}_{l}]}{\sqrt{\sum\nolimits_{l=1}^{\hat{L}}\Vert\hat{\textbf{Q}}_{l}\hat{\textbf{h}}_{\mathrm{DL}}[\hat{k}_{l}]\Vert^{2}}},
\end{equation}
and
\begin{equation}\label{eq:24}
\hat{\textbf{f}}^{\mathrm{MRT}}_{l}=\frac{\sqrt{P_{\mathrm{DL}}}\hat{\textbf{h}}_{\mathrm{DL}}[\hat{k}_{l}]}{\Vert\hat{\textbf{h}}_{\Sigma}\Vert},
\end{equation}
respectively, where  $\hat{\textbf{Q}}_{l}=\textbf{I}_{M}-\hat{\textbf{H}}_{l}(\hat{\textbf{H}}_{l}^{H}\hat{\textbf{H}}_{l})^{-1}\hat{\textbf{H}}_{l}^{H}$ with $\hat{\textbf{H}}_{l}=[\hat{\textbf{h}}_{\mathrm{DL}}[\hat{k}_{1}],...\hat{\textbf{h}}_{\mathrm{DL}}[\hat{k}_{l-1}],\hat{\textbf{h}}_{\mathrm{DL}}[\hat{k}_{l+1}],...,\hat{\textbf{h}}_{\mathrm{DL}}[\hat{k}_{\hat{L}}]]$, and $\hat{\textbf{h}}_{\Sigma}=[\hat{h}_{\mathrm{DL}}^{T}[\hat{k}_{1}],...,\hat{h}_{\mathrm{DL}}^{T}[\hat{k}_{\hat{L}}]]^{T}$.
Also, the estimated effective channel $\hat{\textbf{g}}^{H}_{l}[i], i\in\{\pm1,\cdots,\pm(K-1)\}$ can be obtained similar as \eqref{eq:8}, and the MMSE beamforming vectors are
\begin{equation}\label{eq:25}
\hat{\textbf{f}}_{\Sigma}^{\mathrm{MMSE}}=\sqrt{P_{\mathrm{DL}}}\frac{\hat{\textbf{C}}^{-1}\hat{\textbf{h}}_{\Sigma}}{\Vert\hat{\textbf{C}}^{-1}\hat{\textbf{h}}_{\Sigma}\Vert},
\end{equation}
where $\hat{\textbf{C}}=\sum\nolimits_{i=-(K-1), i\neq0}^{K-1}\hat{\textbf{g}}_{\Sigma}[i]\hat{\textbf{g}}_{\Sigma}^{H}[i]+\frac{\sigma^{2}}{P_{\mathrm{DL}}}\textbf{I}$ with $\hat{\textbf{g}}_{\Sigma}[i]=[\hat{\textbf{g}}_{1}^{T}[i], \cdots, \hat{\textbf{g}}_{L^{\prime}}^{T}[i]]^{T}$.
\par
Next, we derive the resulting SINR for the above DAM designs based on estimated channel. For any designed path-based beamforming vecotrs $\{\hat{\textbf{f}}_{l}\}_{l=1}^{\hat{L}}$ and delay pre-compensations $\{\hat{\kappa}_{l}\}_{l=1}^{\hat{L}}$, the received signal at the UE is
\begin{equation}\label{eq:26}
y[n]=\sum\nolimits_{k=0}^{K-1}\sum\nolimits_{l=1}^{\hat{L}}\textbf{h}_{\mathrm{DL}}^{H}[k]\hat{\textbf{f}}_{l}s[n-(\hat{\kappa}_{l}+k)]+z[n]
\end{equation}
To derive the resulting SINR of \eqref{eq:26}, similar to \eqref{eq:9}, it is necessary to group those symbols with equal delays. To this end, we define the following set
\begin{equation}\label{eq:27}
\Omega_{j}=\{(l,k), 1\leq l \leq \hat{L}, 0 \leq k \leq K-1|\hat{\kappa}_{l}+k=j\}
\end{equation}
where the delay $j$ ranges from $j_{\mathrm{min}}=\min\limits_{1\leq l \leq \hat{L}, 0 \leq k \leq K-1}(\hat{\kappa}_{l}+k)$ to $j_{\mathrm{max}}=\max\limits_{1\leq l \leq \hat{L}, 0 \leq k \leq K-1}(\hat{\kappa}_{l}+k)$.
Therefore, \eqref{eq:26} can be equivalently expressed as
\begin{equation}\label{eq:28}
y[n]=\sum\nolimits_{j=j_{\mathrm{min}}}^{j_{\mathrm{max}}}\Big(\sum\nolimits_{(l,k)\in\Omega_{j}}\textbf{h}_{\mathrm{DL}}^{H}[k]\hat{\textbf{f}}_{l}\Big)s[n-j]+z[n].
\end{equation}
By assuming that the UE locks to the strongest path, the resulting SINR is
\begin{equation}\label{eq:29}
\gamma=\frac{\Big|\sum\nolimits_{(l,k)\in\Omega_{j^{\ast}}}\textbf{h}_{\mathrm{DL}}^{H}[k]\hat{\textbf{f}}_{l}\Big|^{2}}
{\sum\nolimits_{j=j_{\mathrm{min}},j\neq j^{\ast}}^{j_{\mathrm{max}}}\Big|\sum\nolimits_{(l,l^{\prime})\in\Omega_{j}}\textbf{h}_{\mathrm{DL}}^{H}[k]\hat{\textbf{f}}_{l}\Big|^{2}+\sigma^{2}},
\end{equation}
where
\begin{equation}\label{eq:30}
j^{\ast}=\mathop{\arg\max}_{j_{\mathrm{min}}\leq j \leq j_{\mathrm{max}}}\Big|\sum_{(l,k)\in\Omega_{j}}\textbf{h}_{\mathrm{DL}}^{H}[k]\hat{\textbf{f}}_{l}\Big|^{2}.
\end{equation}\par
Let $n_{c}$ denote the number of symbol durations for each channel coherence block, and $n_{g}$ denote the length of guard interval to avoid inter-block interference for DAM. Based on \eqref{eq:29}, the achievable rate for DAM with estimated channel by taking into account the training overhead is
\begin{equation}\label{eq:31}
R = \frac{n_{c}-n_{g}-N}{n_{c}}\log_{2}(1+\gamma),
\end{equation}
where $N$ denotes the number of training intervals.
\section{Simulation Results}
In this section, simulation results are presented to evaluate the performance of the BOMP-based channel estimation method and the resulting performance for DAM. We assume that the BS is equipped with $M=64$ antennas, the carrier frequency is $f=28$ GHz, the sampling interval is $T_{s}=10$ ns, and the noise power is $-94$ dBm. The raised consine pulse shaping function with a roll-off factor $0.5$ is used for $p(t)$. The transmit power in uplink is $P_{\mathrm{UL}}=26$ dBm. The channel coherence time is $T_{c}=1$ ms. The total number of signal samples and the length of guard intervals within each coherent block are $n_{c}=10^{5}$ and $n_{g}=100$, respectively. The total number of channel taps is $K=50$ and the number of multi-paths is $L=5$. The threshold for selecting significant channel taps is $C=0.01$. All the following results are averaged over $10^{3}$ channel realizations.\par
\begin{figure}[h]
\centering
\includegraphics[width=6.9cm,height=4.8cm]{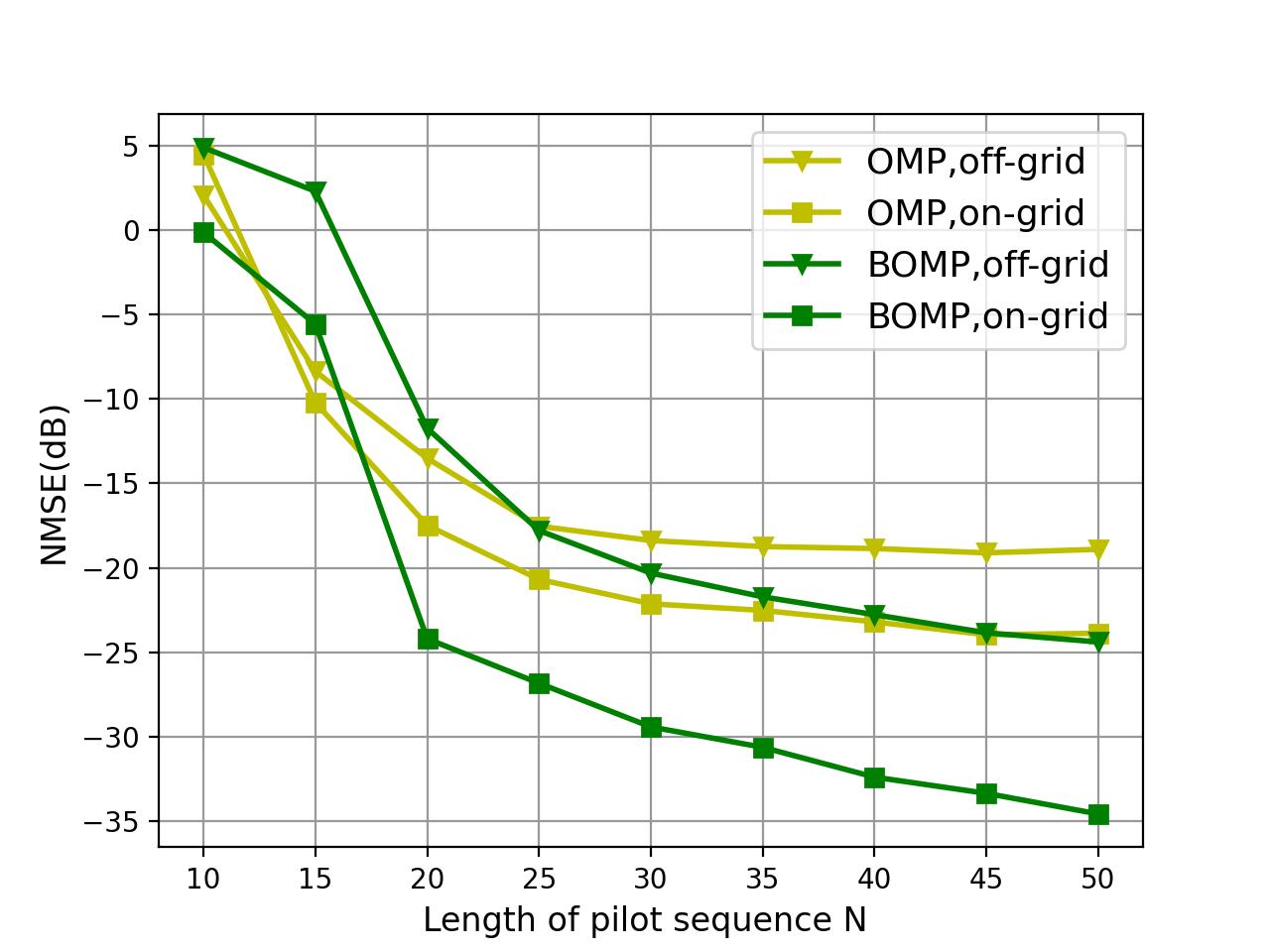}
\caption{The NMSE versus pilot length $N$.}
\end{figure}
For evaluating the accuracy of the channel estimation method in Algorithm 1, the normalized mean square error (NMSE) is considered, which is defined as
$\mathrm{NMSE}=\Vert\hat{\textbf{d}}-\textbf{d}\Vert^{2}/\Vert\textbf{d}\Vert^{2}$.
The training sequence $x[n], n=0,1,...,N-1$ consists of equipropable binary phase shift keying (BPSK) symbols. Furthermore, the OMP-based channel estimation is considered as a benchmark \cite{19}. Both on-grid and off-grid delays are considered. For the former, the path delays are integer multiples of $T_s$, whereas for the latter, they are continuous values in $[0,(K-1)T_{s}]$. Fig. 1 plots the NMSE as a function of the pilot length $N$.
It is observed that as $N$ increases, the NMSE decreases for all schemes, as expected. Besides, for both OMP and BOMP methods, the channels with on-grid path delays can be better estimated, since they exhibit more sparsity. Besides, the BOMP-based estimation outperforms the OMP-based method, thanks to its exploitation of block sparsity.\par
\begin{figure}[h]
\centering
\includegraphics[width=7.8cm,height=5.5cm]{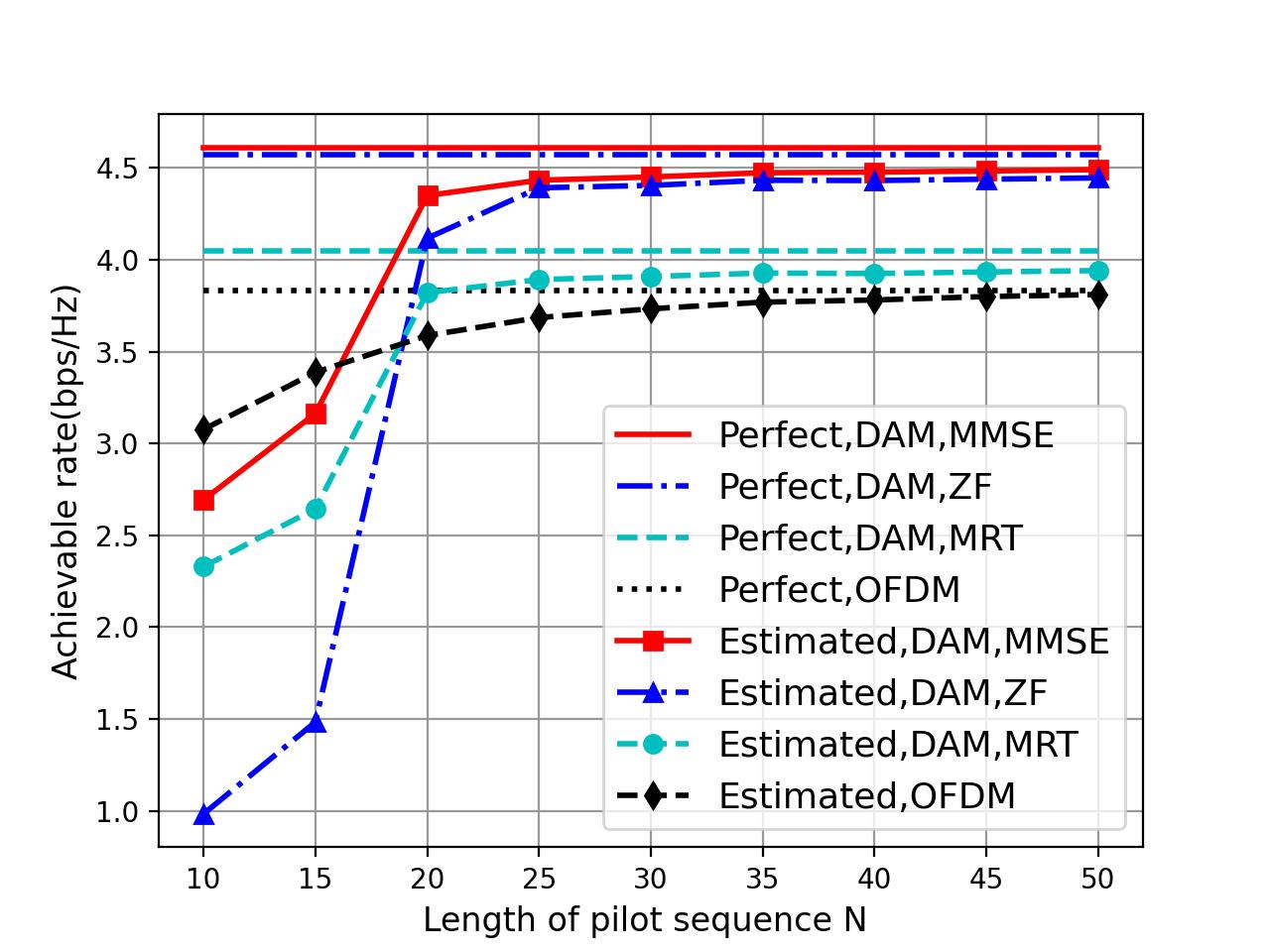}
\caption{The achievable rate versus pilot length $N$ for OFDM and DAM.}
\end{figure}
To evaluate the performance of DAM, the benchmarking OFDM schemes based on perfect and estimated CSIs are also considered, where the channel estimation for OFDM is also based on the BOMP algorithm with uplink training \cite{20}. The number of OFDM sub-carriers is $S = 512$, and a cyclic prefix (CP) of length $N_{\mathrm{cp}} = 50$ is used. The total CP overhead of OFDM for each coherent block is $n_{c}/(S+N_{\mathrm{cp}})\times N_{\mathrm{cp}}$, which is much larger than that is required by DAM ($n_{g}=100$). The pilot length $N$ for the OFDM scheme represents the  number of pilot sub-carriers in a OFDM symbol, and the pilot sub-carriers are equispaced with maximum distance \cite{21}. Besides, the optimal sub-carrier-based MRT beamforming and the classic water-filling (WF) power allocation are performed for the OFDM.
Fig. 2 shows the achievable rate for OFDM and DAM with different beamforming schemes versus the pilot length. The transmit power is set as $P_{\mathrm{DL}}=30$ dBm. It is observed that with perfect CSI, all the three DAM beamforming schemes outperform the OFDM scheme, thanks to the saving of guard interval overhead \cite{1}. For all the three beamforming schemes of DAM, as the pilot length $N$ increases, their performance based on the estimated channel approaches to the ideal case with perfect CSI, which is in accordance with the channel estimation results in Fig. 1. It is also observed that when the pilot length $N\geq 20$, the DAM scheme outperforms OFDM under imperfect CSI, which demonstrates the effectiveness of DAM transmission and the BOMP-based channel estimation.\par
\begin{figure}[h]
\centering
\includegraphics[width=6.8cm,height=5.2cm]{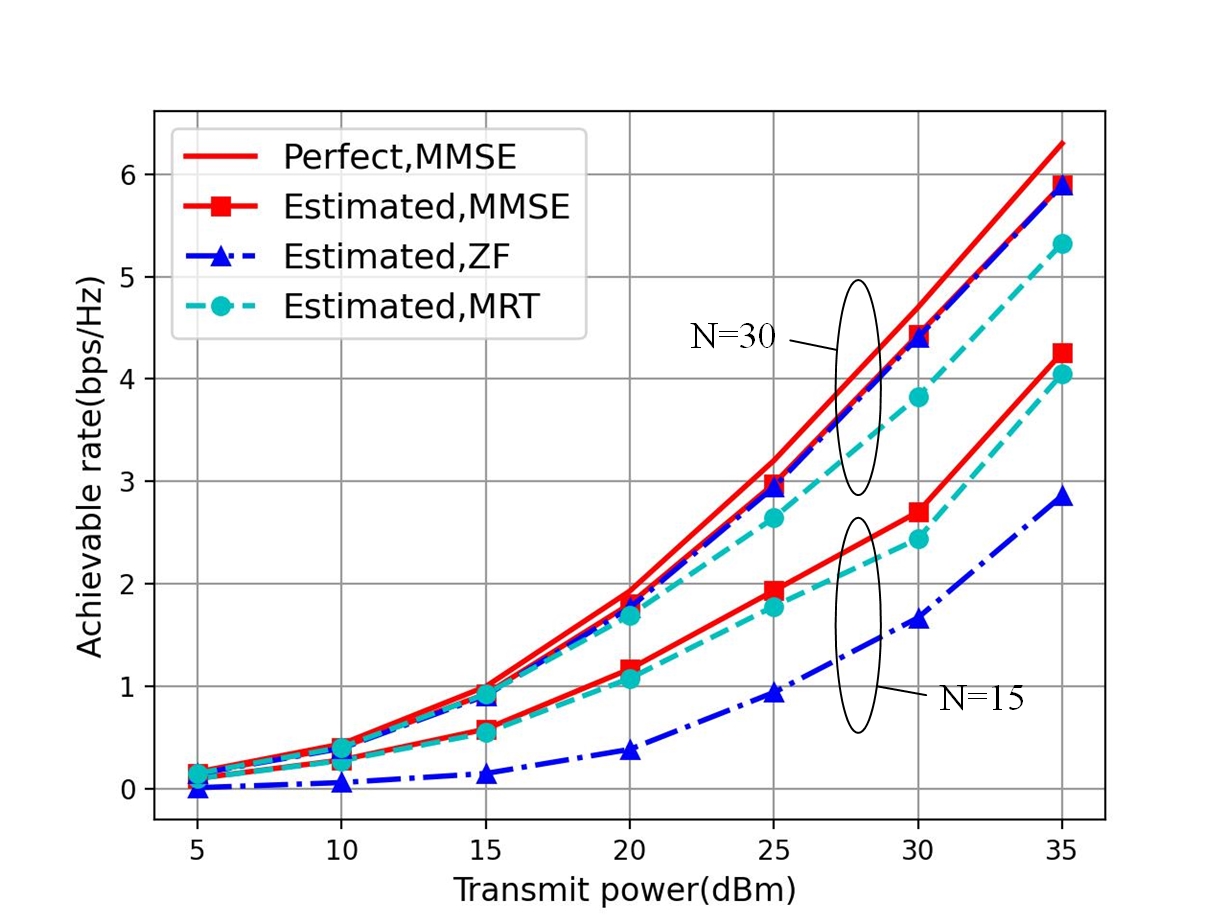}
\caption{The achievable rate of various DAM beamforming schemes versus transmit power $P_{\mathrm{DL}}$.}
\end{figure}
Fig. 3 shows the achievable rate for different DAM beamforming schemes versus transmit power $P_{\mathrm{DL}}$, with pilot length $N=15$ and $30$. The upper bound achieved with MMSE beamforming scheme based on perfect CSI is also plotted. As the transmit power increases, the achievable rates of all schemes increase, as expected. When $N=15$, there is a large performance gap for DAM based on estimated CSI and that based on perfect CSI, while such performance gap reduces when $N=30$, thanks to the accurate channel estimation with Algorithm 1.\par
\section{Conclusion}
In this letter, an efficient BOMP-based channel estimation scheme was studied for DAM. By exploiting the block sparsity of the channel vector, the BOMP-based channel estimation scheme can accurately recover the CSI with low training overhead. Furthermore, the DAM delay pre-compensations and tap-based beamforming based on the estimated CSI were studied. Simulation results demonstrated that with a low training overhead, accurate CSI estimation can be achieved, and the DAM design based on the estimated CSI gives comparable performance as the ideal case with perfect CSI, which also outperforms OFDM due to the saving of guard interval overhead.\par

\end{document}